# The ghost of ecology in chaos: combining intransitive and higher order effects

John Vandermeer [1,2] and Ivette Perfecto [2]

1. Department of Ecology and Evolutionary Biology
   University of Michigan, Ann Arbor, MI 48109
2. School of Environment and Sustainability
   University of Michigan, Ann Arbor, MI 48109

**Abstract**

Historically, musings about the structure of ecological communities has revolved around the structure of pairwise interactions, competition, predation, mutualism, etc. . . Recently a growing literature acknowledges that the baseline assumption that the "pair" of species is not necessarily the metaphorical molecule of community ecology, and that certain structures containing three or more species may not be usefully divisible into pairwise components. Two examples are intransitive competition (species A dominates species B dominates species C dominates species A), and nonlinear higher-order effects. While these two processes have been discussed extensively, the explicit analysis of how the two of them behave when simultaneously part of the same dynamic system has not yet appeared in the literature. A concrete situation exists on coffee farms in Puerto Rico in which three ant species, at least on some farms, form an intransitive competitive triplet, and that triplet is strongly influenced, nonlinearly, by a fly parasitoid that modifies the competitive ability of one of the species in the triplet. Using this arrangement as a template we explore the dynamical consequences with a simple ODE model. Results are complicated and include include alternative periodic and chaotic attractors. The qualitative structures of those complications, however, may be retrieved easily from a reflection on the basic natural history of the system.

Historically, theoretical community ecology has centered on the issue of how many species a community can contain in a stable state. Early literature sought indices, such as the determinant of the community matrix (Levins, 1968) or its eigenvalues (May, 1971) as generalized measurements of stability. Today, such classical notions from dynamic systems theory are generally eschewed and a body of sophisticated theory has evolved elaborating new approaches to the classical theme, remaining within the generalized framework of stability-like questions associated, broadly speaking, with biodiversity (Simha et al., 2022). A large and growing literature seeks to understand these ideas partially through the vehicle of per capita population increase of a new species facing an otherwise established community, the "invasion growth rate" (MacArthur and Levins, 1967; Barabás et al, 2018; Chesson, 2000), under the popular theme of "modern coexistence theory" (Ellner et al., 2019; Hofbauer and Schreiber, 2022 ).

Perplexing issues are readily acknowledged (Barabás et al., 2018; Allesina and Levine, 2011), especially associated with failure to recognize candidates for inclusion as perhaps belonging to a structure of higher dimension, in which two or more species unable to enter the community individually, are permitted entry through their joint dynamics. That is, a given population may be unable to join a particular assemblage of species alone (i.e., have a negative or zero invasion growth rate), but in the context of other interactions with other species, might easily do so – e.g., obligate mutualists (Johnson, 2021) or species with a significant Allee effect (Godwin et al., 2020). Recent literature has noted two particularly obvious candidate dynamical structures that challenge the one-dimensional marginal per capita growth framing – intransitive structures and higher order effects (Levine et al., 2017; Barabás et. al., 2018).

Intransitive competition (as in the game rock-paper-scissors) emerges naturally from elementary musings about the process of competition (May and Leonard, 1975) wherein, for example, plant species A may competitively dominate species B which dominates species C in light competition whereas C may dominate A in competition for underground nutrients, leading to the potential intransitive competition of A>B>C>A. Such arrangements are increasingly recognized as potentially important in community structure (Soliveres et al., 2015; Gallien et al., 2017), even though demonstrating them in nature remains challenging (Allesina and Levine, 2011).

Similarly, higher order effects generally require a group of species within which a low dimensional subset (perhaps only two species) has its interaction compromised or encouraged by a third species. For example, the competitive dominance of one species of amphibian over another in Michigan ponds is compromised by nothing more than the odor of an odonate predator (Peacor and Werner, 2001). That is, the actual density-mediated predatory effect (odonate eats frog) may be trivial compared to the indirect action on the trait of competition, by means of the very presence of the predator (frog smells predator and hides, leaving resources available for its competitor). The importance of these higher order effects has been long appreciated (Vandermeer, 1969; Polis, 1991) and recently gained prominence theoretically (Bairey et al., 2016; Angulo et al., 2021), complementing a substantial number of empirical examples (Werner and Peacor, 2003; Hsieh and Perfecto, 2012; Hsieh et al., 2012; Offenberg, 2015; Schifani et al., 2020; Barbosa et al., 2023).

Significant complications arise with the consideration of both intransitive structures and higher order effects acting together. Intransitivities can take many forms and insert themselves in species assemblages in quite complicated ways (Vandermeer, 1988; Vandermeer and Jackson, 2018; Vandermeer and Yitbarek, 2012), perhaps even providing a core structure for larger communities (Vandermeer and Perfecto, 2023). Similarly, higher order effects may take on a variety of forms, indeed, the very cataloging of such effects is challenging (Battison et al., 2020).

Nevertheless, a simple question may be posed – what might be the consequences of the simultaneous action of intransitivity and higher order effects?

Here we propose a specific framework for studying one version of this question, using an exemplary system now well-known to operate at least in some situations. A group of ant species seemingly ubiquitous on coffee farms in Puerto Rico, has been under study for the past 6 years and arguably provides an example of a community that is dominated by an intransitive competitive structure modified in a non-linear way by a higher-order predator prey system connected to it (Perfecto and Vandermeer, 2020; Vandermeer and Perfecto 2020; 2023). Since the intransitive structure is, by its nature, oscillatory, the addition of a predator prey oscillation represents an ecologically unique focus on coupled oscillators in ecosystems, and the indirect, trait-mediated effect of the predator adds yet another higher order effect. We present a series of quantitative field observations and experiments supporting the hypothesis that a major organizing metaphor for this community is only understandable if we acknowledge that an intransitive competitive structure intersects a higher order effect, the predator/prey interaction induced by the phorid. In what follows we first describe the natural history of the system, then develop some simple theoretical framings in the spirit of a toy model or "thought experiment," elaborating on the fact that an intransitive loop is inherently coupled to a higher order effect associated with a predator/prey situation (Vandermeer and Perfecto, 2023).

**Natural history of the model system**

Previous research (Perfecto and Vandermeer, 2020) established a varied structure of the community of arboreally foraging ants on a group of 25 coffee farms throughout the coffee-growing region of Puerto Rico. Farm-to-farm variability is evident and the "coffee farm ant community" is as good a "metacommunity" as any we have heard of. Through the variability of the system, three particular species seemed quite common, *Solenopsis invicta, Wasmannia auropunctata*, and *Monomorium floricola*, all non-native species and hereafter referred to by their generic identities. Subsequent research (Vandermeer and Perfecto, 2020) argued that the competitive effect of *Solenopsis* on *Wasmannia* was relatively large on one particular farm, and suggested that the modifying effect of a parasitic fly (*Pseudacteon* spp, Phoridae) modified that competitive effect, qualitatively implying an indirect higher order effect (the phorid affecting the strength of the *Solenopsis* competitive effect on *Wasmannia*), reflecting a large literature on predator-mediated coexistence (Caswell, 1978; Schreiber, 1997; Hsu, 1981). Recently it has been reported that these three ant species form an intransitive competitive loop, with the local replacement series of *Solenopsis* replaces *Wasmannia,* which replaces *Monomorium*, which replaces *Solenopsis*. Furthermore, abundant casual observations on coffee farms throughout Puerto Rico suggest that the phorid flies are abundant throughout the coffee-growing region. The basic structure of this natural history is presented in figure 1.

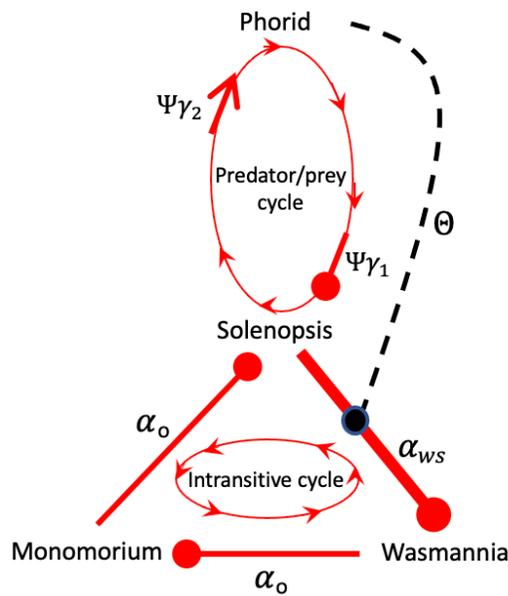

*Figure 1. Diagrammatic representation of the system, emphasizing its coupled oscillator nature, in the form of a directed hypergraph, with parameters of the equation system, to be stated later, indicated on the edges. Filled circles indicate a negative effect, larger arrowhead indicates positive effect (small arrowheads generally indicate oscillations). The phorid and Solenopsis pair is a predator prey oscillator and the three ants, Solenopsis, Wasmannia, and Monmorium form an intransitive loop, also an oscillator.*

The details of this system have been studied on one particular farm (UTUA2 in Perfecto and Vandermeer, 2020 – see also supplementary material herein) on a plot of 150x60 m for the past 5 years (Vandermeer and Perfecto, 2023). The habitat consists of an intercrop of coffee and citrus trees (Fig. 2), with distinct potential niches for the ants being 1) citrus trees, 2) coffee bushes, and 3) terrestrial (including subterranean foraging trails).

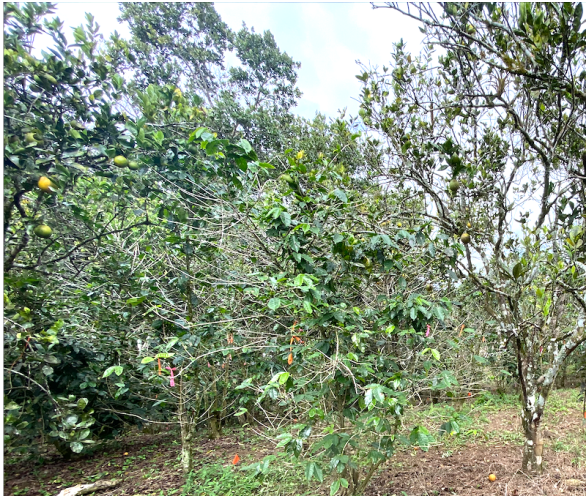

*Figure 2. General appearance of the farm on which these studies took place. Note the coffee bush in the foreground and the citrus tree to the right. The ground foraging area ranges from the relatively open area on the right foreground to more dense understory as on the left.*

Elsewhere we have argued for the intransitive nature of the three most common species, *Solenopsis, Wasmannia* and *Monomorium*, including a spatially explicit time series reflecting the intransitive dynamics (Vandermeer and Perfecto, 2023). Repeat sampling over a 12-month period on coffee bushes, and citrus trees, and transects on the ground twice a year, provide insights on the natural history of the system. Key to understanding the behavior of the system within the framework presented herein, is the non-random nature of the phorid attacks on *Solenopsis*. In figure 3a we illustrate the distribution of phorid sightings from three independent censuses of phorids at *Solenopsis* nest mounds on the long-term census plot (see Vandermeer and Perfecto, 2023 for details). It is evident that approximately the eastern half of the plot is where the overwhelming majority of phorid attacks were encountered, fitting nicely with the "predator-mediated" control over the particular competitive interaction between *Solenopsis* and *Wasmannia* (Vandermeer and Perfecto, 2020; 2023), allowing *Wasmannia* to enter the system even in the presence of *Solenopsis*. In the western half of the plot, where the phorid parasitoids were effectively absent, the takeover of *Wasmannia* territory by *Solenopsis* (especially on the ground) proceeded as anticipated in late 2022 and early 2023 (Vandermeer and Perfecto, 2023).

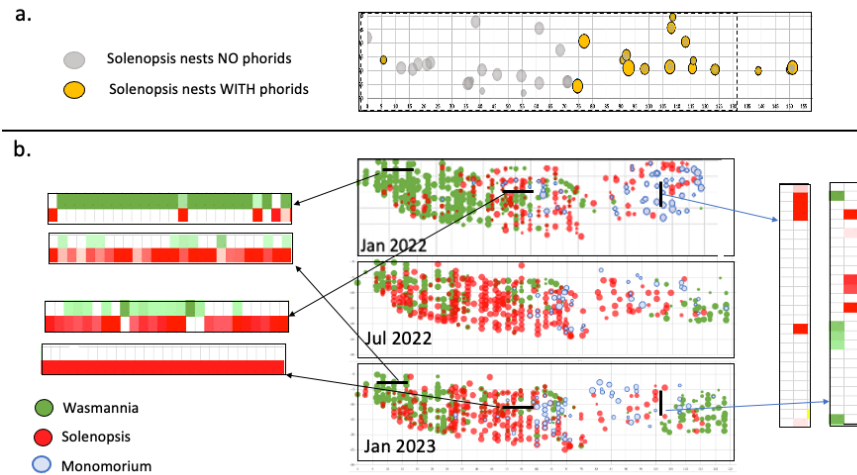

F

*Figure 3. a. Presence (darker yellow) or absence (lighter grey) of phorid flies on Solenopsis nest mounds on the main plot, with size of the darker yellow bubble representing the number of times (from 1 to 3) that phorids were observed and the size of the less dark grey bubbles representing the number of times (from 1 to 3, sampled in January, 2023) the nest was observed to lack phorids over a period of a 2 minute observation. b. Ground sampling results at three sites on the plots. Center three graphs are of the coffee trees and the Wasmannia (green), Solenopsis (red) and Monomorium (blue) over the course of a single year at 6 month intervals. Linear graphs to the left and right are ground samples taken in January of both years at the same precise locations. Note how the left-hand graphs illustrate the takeover of Solenopsis on the ground while the ones on the right-hand side show the beginnings of the invasion of Wasmannia into the area, correlated with the presence of the phorid parasitoid, as illustrated in part a.*

There is also a dramatic difference in the ecological niches of these three species (Vandermeer et al., 2022; Vandermeer and Perfecto, 2023), basically supporting the general intransitive structure. In particular, the foraging area of *Solenopsis* extends in underground tunnels approximately 10 meters in all directions from the evident above-ground nest mound. The tunnels are interrupted periodically by foraging exit holes. It has been argued that this particular niche construction contributes significantly to the competitive ability of *Solenopsis*. Repeated ground sampling (using tunafish baits) reveals an on-the-ground and below-ground dynamic that corresponds to the takeover by *Solenopsis* of coffee bushes originally dominated by *Wasmannia* through a gradual dominance of the ground area below, as seen in the western half of the plot in figure 3b (also see Vandermeer and Perfecto, 2023). Contrarily, *Wasmannia* seems to be in the process of taking over coffee bushes previously dominated by *Solenopsis* as seen at the eastern end of the plot in figure 3b, through either or both of 1) the direct competitive effect of the third species, *Monomorium*, in the intransitive triplet, and/or 2) the pressure exerted by the phorids in the eastern half of the plot which reduces the *Solenopsis* competitive effect on the *Wasmannia*, perhaps even reversing the dominance (Fig. 3b, also see Vandermeer and Perfecto, 2020).

Finally, an important niche distinction is the apparent inability of *Monomorium* to forage on the ground, being restricted to coffee bushes and citrus trees in this example. The general dominance of the three species foraging on citrus trees over a 6-month period is presented in figure 4, along with a vertically compressed plot of the coffee bush occupancies (from figure 3b), reflecting the dynamics evident in the coffee bushes. In the western half of the plot, the *Solenopsis* took over many of the citrus trees between Jan 2022 and 2023, but there is a major increase of the *Wasmannia* in the eastern half of the plot over that same time period, presumably due to the competitive dominance of the latter over *Monomorium*.

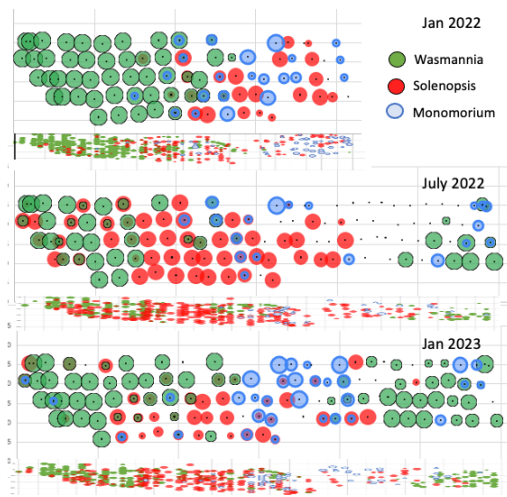

*Figure 4. Occupation of citrus trees over a 12 month cycle. Circle sizes correspond to the number of baits occupied on a given citrus tree (from 0 – 5) and the compressed graphs below each plot is compressed from the data in figure 3b, illustrating the strong correlations between citrus occupancy and coffee occupancy.*

All of this natural history supports 1) the existence of an intransitive loop that consists of *Solenopsis>Wasmannia>Monomorium>Solenopsis*, probably discernable in practice only in a spatial context (see also Vandermeer and Perfecto, 2023), and that 2) a significant force connecting with this triad is a predator prey system that involves a higher order effect. Importantly, both of these non-pairwise-interactions generate oscillations and occur simultaneously, raising the theoretical question of what are the consequences of such coupled non-linear oscillators, a point to which we turn presently.

**Theoretical dynamics of an intransitive loop coupled with a predator/prey cycle**

Although both intransitive structures (Soliveres and Allan, 2018; Allesina et al) and higher order effects (Batteson et al., 2020) contain idiosyncrasies that make generalizations difficult, a simple abstraction tied to this real-life example, in the spirit of a "toy model" or "thought experiment" may be useful. We propose a set of ODEs in such a spirit, with the four variables S for *Solenopsis*, P for phorids, M for *Monomorium* and W for *Wasmannia*, as follows:

$$\frac{dS}{dt} = S(1 - S - \alpha_0 M - \psi\gamma_1 P) \quad \text{1a}$$

$$\frac{dP}{dt} = \psi\gamma_2 KPS - mP \quad \text{1b}$$

$$\frac{dM}{dt} = rM(1 - M - \alpha_0 W) \quad \text{1c}$$

$$\frac{dW}{dt} = W(1 - W - \alpha_{ws}\theta S) \quad \text{1d}$$

$$\psi = \frac{1}{1 + fS} \quad \text{1e}$$

$$\Theta = \frac{1}{1 + \beta P} \quad \text{1f}$$

where $\alpha_i$ is the competition coefficient in the intransitive loop, $\gamma_i$ are the predatory consumption and conversion rates, m is the predatory mortality rate, β is the indirect trait-mediated interaction term, and f is the functional response term (reflecting "handling time" and/or predator satiation).

The equations are diagrammed in figure 1. In the absence of the phorid parasitoid the dynamics of the intransitive triplet are well-known-- if the competition coefficients are all identical and > 2.0, the system oscillates in a heteroclinic cycle focused on the three equilibria, {1, 0, 0}, {0, 1, 0}, {0, 0, 1} , repeatedly visiting the neighborhood of each of the points, ever closer to the actual values but, theoretically, never actually attaining their values (May and Leonard, 1975; Zeeman, 1993). We refer to such a situation as an unstable equilibrium (formally, a heteroclinic cycle). If the competition coefficients are all < 2.0, the system exhibits damped oscillations to a stable equilibrium, where each value is 1/3 (provided that $\alpha_{ws} = \alpha_o$). In either case the system is oscillatory, dissipative if $\alpha < 2$, heteroclinic if $\alpha > 2$.

Adding the phorid to the system generates two higher order effects. First, in the tradition of predator prey systems representing classical oscillators, the phorid/*Solenopsis* system combines with the intransitive triad, also an oscillator, to create a system of coupled oscillators, effectively adding a higher order component into the system. Second, as is well-known with phorids and ants (Hsieh and Perfecto, 2012; Morrison et al., 1999; 2000) the phorid elicits a behavioral response in the ants, such that the ant's vital rates (including its competitive effect on others) may be moderated, a second class of higher order effect. These dynamic elements are illustrated in figure 1, wherein $\alpha$ is the competition coefficient of the intransitive competitive triad, $\gamma_1$ is the predation rate, $\gamma_2$ is the consumption rate, $\Psi$ is the functional response term for the predator/prey system, and $\Theta$ is the non-linear trait-mediated indirect effect of the phorid on the ability of *Solenopsis* to compete with *Wasmannia*. The functional meaning of $\Psi$ and $\Theta$ are stipulated in equations 1e and 1f.

We begin with a symmetric version of the system where $\alpha_0 = \alpha_{ws}$, and $\beta = 0$. Extensive numerical study of this system reveals several rather mundane generalizations, presented in detail in the supplementary material (S1). Furthermore, simply tracing the qualitative effects on the graph in figure 1 allows a certain degree of prediction. We summarize these predictions with reference to the network diagram in figure 1, and later employ equation system 1 to suggest certain restrictions on possible outcomes. There are six species extinction scenarios that can be discerned from a general look at the graph (Fig. 1), and result from employment of the symmetrical version of systemn 1: 1) All species may persist, although the details of that persistence may be somewhat complicated; 2) The phorid may be too strong a predator and eliminate the *Solenopsis* from the system, subsequently dying of starvation, while the *Monomorium* will succumb to the *Wasmannia* and the latter persisting alone; 3) The *Solenopsis* will eliminate the *Wasmannia* and then succumb to competitive elimination from the *Monomorium* while the phorid dies of starvation, leaving a single species, *Monomorium*; 4) the *Wasmannia* eliminates the *Monomorium* taking pressure off of the *Solenopsis* which consequently can eliminate the *Wasmannia*, leaving the basic predator prey system of phorid and *Solenopsis* as a persistent oscillatory system; 5) the *Wasmannia* similarly eliminates the *Monomorium*, but the *Solenopsis*/phorid system remains leaving the three species, phorid/*Solenopsis*/*Wasmannia*; and 6) the basic predator/prey system is unstable and the phorid dies, leaving the original intransitive loop of *Solenopsis*/*Wasmannia*/*Monomorium* triad. Any of these scenarios is possible and completely depends on the values of the parameters in the system and the initial conditions, which are rather difficult to estimate in the field (but see Vandermeer and Perfecto, 2020). On the other hand, a variety of scenarios are impossible based on the qualitative nature of the network. For example, other than the phorid/*Solenopsis* predator/prey system alone, there are no possible outcomes with only two species surviving.

Extensive numerical solutions of equation system 1 reveal an additional qualitative pattern (summarized in table S1 in the supplementary material). Beginning with a "super stable" intransitive loop ($\alpha = 0.9$), and the predation system in a focal point, coupling the oscillators leads to a four-species focal point (Fig. S2). Increasing the competition coefficient (to 1.9 -- still in its stable form) results in the four species system decaying to a three species focal point with *Monomorium* going extinct (Fig. S3). However, if the initial system includes the predation system initially in a limit cycle, the intransitive competitive system can break down as the stability of the intransitive loop is decreased (as "$\alpha$" moves toward the 2.0 threshold) and a heteroclinic cycle emerges driving one of the competitors to extinction (most frequently *Monomorium*) (Fig. S4).

In contrast, if we begin with the intransitive competitive system in a heteroclinic cycle, there does not seem to be a scenario in which a stabilizing force is exerted by coupling with the predator/prey cycle, even if the latter is in a focal point structure (Fig. S5). Indeed, if the predator/prey system is in a limit cycle, the trajectories seem to get somewhat erratic, but do not deviate from the basic heteroclinic cycle, usually with *Monomorium* being the first to go extinct.

Relaxing the symmetry assumption for the intransitive loop, a variety of complications emerge, an example of which we now discuss. With the parameter settings as indicated in the caption to figure 5, we see the existence of two very distinct attractors, both oscillatory. The first attractor is a simple oscillation (a 1 point cycle) and the second one is a more complicated cycle with four peaks repeated endlessly (a 4 point cycle). Note, as indicated in the caption to figure 5, that there is only a small difference in initial conditions, yet the alternative attractors are quite distinct. We note in passing that the simple one-point cycle exists, apparently alone, in a slightly different parameter space (i.e., f = 4.6). The four-point cycle is obviously associated with the four state variables wherein each peaks in turn, from *Solenopsis* to *Monomorium* to the phorid to *Wasmannia* and back to *Solenopsis* again, in what is effectively a four species intransitive cycle.

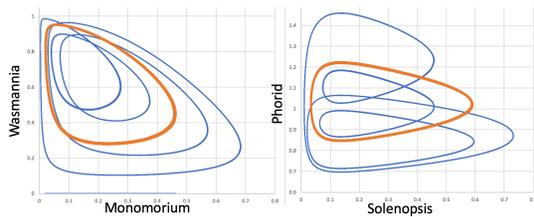

*Figure 5. Alternative attractors in the 4D system (thick red 1-point attractor versus thinner blue 4-point attractor). Parameter values are $\alpha_o = 1.2$, $\alpha_{ws}=5$, $\beta = 2.5$, $\gamma_1=1$, $\gamma_2=0.6$, $f = 4.65$. Initial conditions are Solenopsis = 0.045; Phorid = 1.3; Monomorium = 0.155; Wasmannia = 0.916, for the 1-pt cycle and Solenopsis = 0.045; Phorid = 1.4; Monomorium = 0.155; Wasmannia = 0.916, for the 4-pt cycle.*

Using the parameter of satiation (or handling time, f) as a tuning parameter, if we change it from 4.65 to 5.0 the pattern displayed in figure 6 emerges. Note that (Fig 6a) the trajectories of the two alternative attractors appear similar to the ones pictured in figure 5. However, a microscopic view of either attractor (Fig. 6 b and c) reveals a layered trajectory that is evidently either a quasiperiodic cycle or chaos.

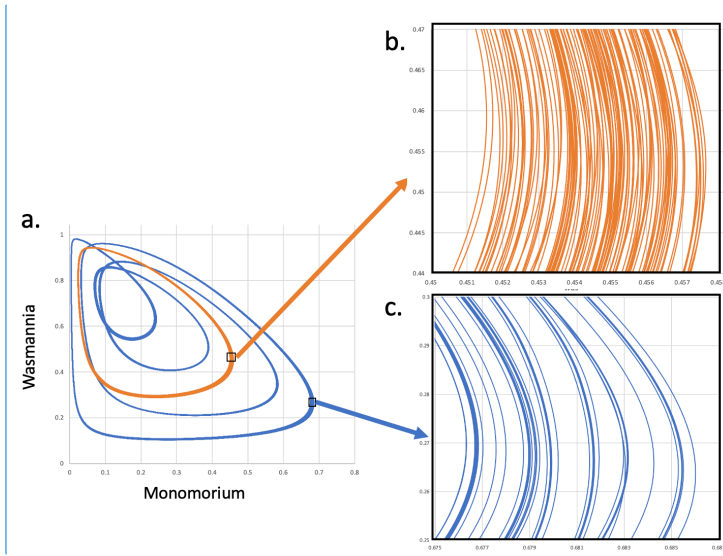

*Figure 6. Alternative attractors in the 4D system (thick red 1-point attractor versus thinner blue 4-point attractor). Parameter values are the same as in figure 5, except f = 5.0.*

A qualitative explanation of both the limit cycle solution (Fig. 5) and the quasiperiodic/chaotic solution (Fig. 6) is suggested by assuming two distinct dynamical structures for the *Solenopsis* variable of the intransitive loop. As in figure 7a, if the effective carrying capacity of the *Solenopsis* is relatively low and the dynamics are heteroclinic, in the absence of phorids two of the three species will go extinct, a standard feature of the heteroclinic framework. If the effective carrying capacity of the *Solenopsis* is relatively high, as in figure 7b, and the dynamics are dissipative (a stable focal point), all three species will survive, oscillating to a single equilibrium point. With the addition of the predator/prey cycle into the system, we expect the *Solenopsis* to range from high density to low density based on the predator prey oscillations, suggesting that the alternative pictures in figures 7a and b will be alternatively visited, perhaps creating a balancing effect and thus a limit cycle, or even present a classic case of such oscillations becoming chaotic, as they evidently do (Fig. 8).

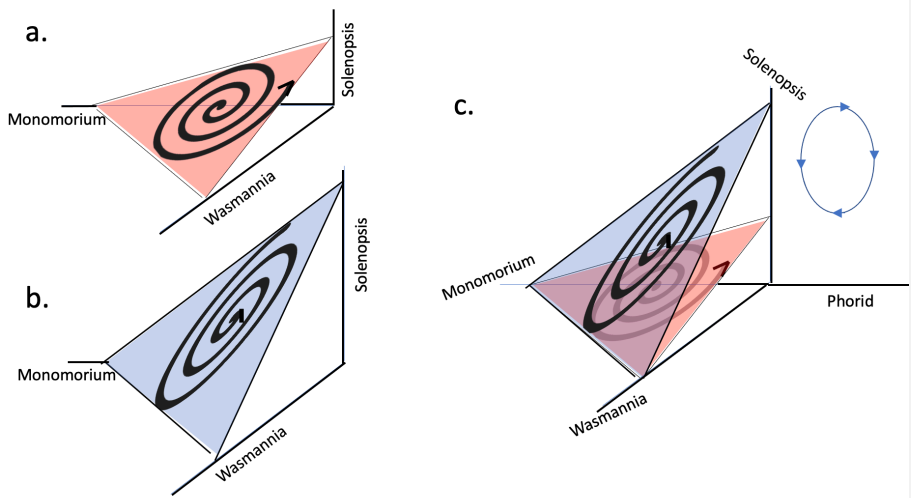

*Figure 7. Expected structure of combining a) a heteroclinic cycle with relatively low population density of* Solenopsis*, with b) a focal point attractor with a relatively high population density of* Solenopsis*. One might imagine how an oscillating population of* Solenopsis *could create a "balance" between the heteroclinic cycle and the focal point. c) When the phorid is in the system, the* Solenopsis *indeed is forced to oscillate, possibly creating an extra complication with these two coupled oscillators. As the Solenopsis oscillates due to its forcing from the phorid, the general manifold of the intransitive triad will move up and down, between a heteroclinic cycle and a focal point, suggesting the qualitative conclusion of a limit cycle, or perhaps chaos or quasiperiodic behavior.*

Under certain parameter settings, the expected (from a qualitative look at figure 7) intersection of a heteroclinic/dissipative cycle (Figs. 7a and b) and the predator prey cycle (Fig 7c) indeed do create an apparently chaotic trajectory, as displayed in figure 8. Yet, within this chaotic structure it is evident that the underlying ecology is visible, sort of a "ghost" appearing from the chaos of the dynamics. With reference to figure 8c, the time series begins with the phorid at its peak, which means the *Solenopsis* is declining, giving rise to the *Wasmannia* increasing to its peak. Competitive pressure of *Wasmannia* on *Monomorium* releases the *Solenopsis* from competition which, combined with a lowered predation from the phorid results in the *Solenopsis* increasing. Increasing *Solenopsis* puts pressure on the *Wasmannia* which relieves pressure on the *Monomorium* which allows it to increase again, adding pressure to the *Solenopsis* which is pressured by the now-increasing phorid. The basic intransitive cycle of the three ant species then dissipates over six cycles, wherein the phorid peaks to start the cycle again, which lasts for about 200 time steps. The qualitative story begun in figure 1 is thus reflected, qualitatively, in the simulations run from the toy model (equation set 1), an "ecological ghost" observable through the mist of formal chaos.

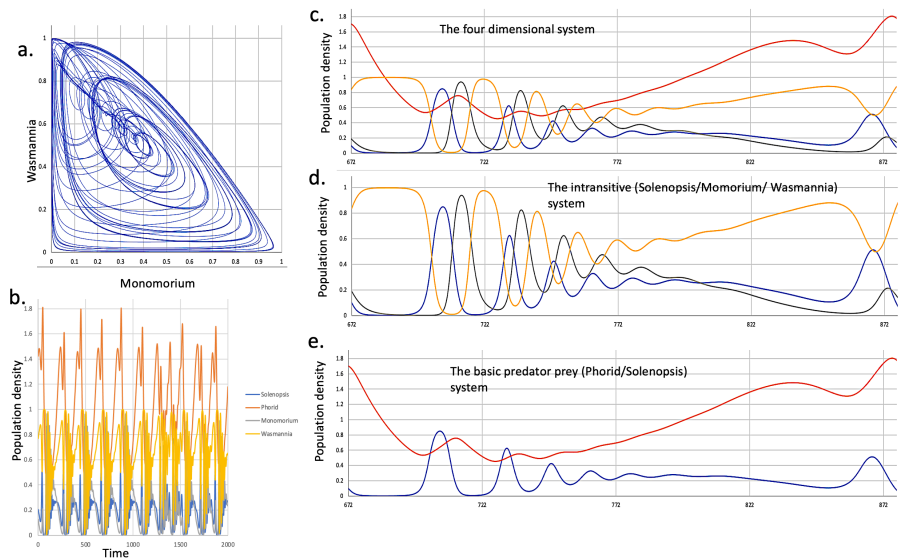

*Figure 8. Details of the chaotic attractor. a. phase space presentation of two of the three intransitive species. b. Time series of the whole system. c. Short time series illustrating the long cycle of predator prey (phorid/Solenopsis) and the faster cycling three species intransitive triplet. d. same as c but with the transitive triplet only. e. same as c but with the predator prey cycle only.*

Finally, parameter space exists in which the system is seemingly chaotic, with the same "ecological ghost," but is in reality a periodic point (Fig. 9), exemplary of the periodic windows frequently seen in chaotic systems. Figure 9a is the same as figure 8a and figure 9b is one cycle from one predator (phorid) peak to the next one (200 times later) in this chaotic attractor. In figure 9c, oscillations are a bit more complicated, yet this is an example of a strictly periodic cycle, not a strange attractor. It is seemingly more "complicated" than the chaotic attractor that emerged in a slightly different parameter space and it has precisely the same ecological ghost, but here there are 13 cycles of the intransitive system for every one cycle of the predator/prey system and the precise trajectory is repeated every 651 time units. Qualitatively, it is the same story, understood as the basic "ecological ghost."

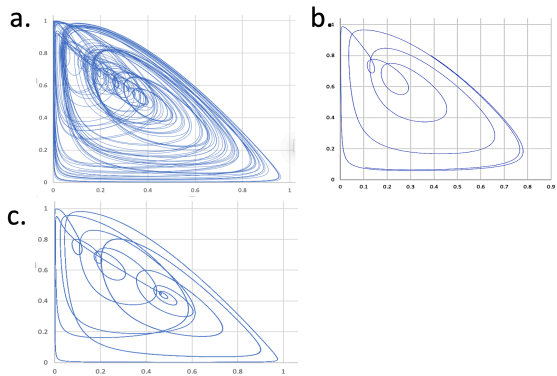

*Figure 9. Phase portraits comparing a) the chaotic attractor and b) one cycle of that chaotic attractor, to c) a periodic "window" of the same system.*

**Discussion**

The two well-known deviations from the pairwise approach to community ecology, intransitivity and higher-order effects, are here merged into an exploration of their joint operation as revealed in a real-world system on coffee farms in Puerto Rico. That the intransitive competition is oscillatory suggests a compelling metaphor where the competitive intransitive oscillation of three species couples with the oscillatory nature of a predator/prey system. Combining real world examples of oscillatory intransitive competition and an evident higher-order interaction introduced by a coupled predator/prey system, we see some complicated and surprising behaviors emerging from a toy model. Furthermore, these complicated structures, some chaotic others not, are evident reflections of the qualitative structure of the original natural system, leading to the metaphor of an ecological ghost underlying the qualitative structure of the model. These two oscillatory frameworks, one from an intransitive competitive triplet the other from a classic predator/prey system, result in a dynamic structure that is, in the end, evident, and could have perhaps been predicted without the mathematical intervention to start with. That is, a basic four cycle system may emerge wherein the predator prey cycle is evident and sets a long term cyclic process, interrupted at more-or-less regular intervals by the more rapid cycling of the intransitive cycle. Reflecting the popular physical model of a double pendulum, a chaotic attractor may emerge, effectively repeating this four cycle pattern, but with sensitive dependence on initial conditions. Perhaps more importantly, periodic points of extreme complexity (but not chaotic) may emerge, following the basic rules of a long term predator prey oscillation within which the heteroclinic cycle oscillates over the relatively short term (e.g, Figs. 8 and 9). This pattern is not, however, inevitable. Indeed, there are situations in which relatively complicated (chaotic or not) alternative attractors can coexist (see figures S1 and S2 in the supplemental material). And, of course, completely unstable formulations exist at many points in the 6-D parameter space, frequently leading to the conclusion that are evident from a simple qualitative glance at the basic structure (Fig 1), and very frequently resulting in the exclusion of *Monomorium*.

Formally, there are two qualitatively distinct parts of parameter space (and consequently dynamic behavior), namely, 1) whether or not the intransitive loop alone presents a stable focus or a heteroclinic cycle (which we sometimes refer to as merely stable versus unstable), and 2) whether the predator prey system alone would present a stable focus or a limit cycle (which we sometimes refer to as merely stable, the focus, or unstable, the limit cycle). As indicated in figure 7, if we imagine two versions of the intransitive loop, one stable the other unstable, we can imagine oscillating between those two states resulting in a constrained periodicity, either a limit cycle or a chaotic attractor. The system we are modelling has the phorid predator causing the *Solenopsis* to oscillate from low to high levels due to its predator/prey dynamics, thus suggesting the chaotic or simply periodic behavior. Simulations with the simple model (equation system 1) show precisely this sort of behavior, including examples in which it could be argued that a non-chaotic periodicity is more "complicated" than a chaotic one (Fig. 9).

The approach taken in this work may be criticized on the general grounds that dynamic structures containing many parameters, may be capable of particularly interesting behavior, yet if that behavior is restricted to a tiny corner of the inevitably multidimensional parameter space with no knowledge of how common that corner might be in nature, we may be searching for the last angel that will fit on the pinhead (Rohani, personal communication). While such a criticism could be made here, we argue that our search for patterns based on generalized structure and comparison of the symmetrical system to the asymmetrical one moves us toward the reality of our framework. Coupling the theoretical machinations with field observations, we feel, further justifies our approach. The "ghost" of the field observations indeed provides a qualitative explanation of what the formal model says.

In addition to the perhaps surprising complex periodic behaviors to be expected from this relatively simple system, much of parameter space suggests simpler dynamics could obtain also, as suggested above by a simple qualitative examination of figure 1. All such results are substantiated by numerical solutions of the basic equations and presented in the supplementary material (see Table S1 for a summary). Notably, many of the qualitatively distinct outcomes of the arrangement result in the extinction of one or more elements. If, as we suggested in the introduction, the coffee ant community is indeed a metacommunity (each farm containing a subcommunity), the dynamic structures we describe herein effectively provide mechanisms for the local extinction half of the metacommunity structure (that structure being 1) long distance dispersal plus 2) local extinction) at a larger scale (i.e., the entire coffee-producing landscape of Puerto Rico). However, the evident results from the elementary equations break down when the symmetry of those equations is relaxed. Parameter modifications that break that symmetry yield exemplary results that are quite complicated, including alternative oscillator attractors, quasiperiodic and chaotic behaviors, and indeed some very complicated oscillations that are strictly periodic and, notably seemingly more "complicated" than the chaotic ones.

An alternative natural instantiation of this system emerged from a detailed study of a different farm, one in which *Monomorium* was not present (Vandermeer and Perfecto, 2020). The intense competition between *Solenopsis* and *Wasmannia* was thus the focus of attention in this study, meaning that, in the present context, the one-way competitive effect of *Solenopsis* against *Wasmannia* was simply one of the outcomes of the model system, and the role of the phorid parasitoid was in the classic vein of predator-mediated coexistence (Schreiber, 1997). Theoretically, it is evident from just a topological examination of figure 1 that if *Monomorium* is not present, the dynamics would be, qualitatively, the same as reported elsewhere (Vandermeer and Perfecto, 2020 – although the focus of that report was embedded in the paradigm of spatial structure). Thus, through the intransitive nature of a part of the overall network, the generally negative effect of the phorids on the *Solenopsis* can be moderated, a complex arrangement that emerges from the particular wiring of the network – an intransitive loop coupled with a nonlinear predator/prey cycle. The predator/prey cycle sets the temporal stage, repeating itself, approximately, every 200 time units, and the smaller oscillations of the intransitive cycle, evidently dissipating, cycle approximately six times for each predator/prey cycle. Thus, the underlying ecology of an intransitive cycle driven by a predator/prey cycle can be easily visualized as a background "ecological ghost" of this chaotic cycle.

An important caveat to this work is the possibility that the scores of other species in this network may themselves form intransitive loops, either alternative triplets or other odd-numbered species combinations (Vandermeer, 2013). Another evident feature is the collection of other ants in the system (Perfecto and Vandermeer, 2020) some of which may be arranged in a transitive hierarchy, but connected to the intransitive loop, an arrangement resulting in a unique form of species coexistence, especially in the context of metapopulations (Vandermeer and Perfecto, 2023). Furthermore, there may exist alternative attractors, each of which is quasiperiodic or chaotic, in the same state space. Finally, each of those attractors may expand in response to a tuning parameter (here we used the parameter of functional response) and seemingly intersect to create a more complicated chaotic attractor.

A chaotic attractor can, despite its infamous complexity and unpredictability, sometimes be intuitively constructed from natural history knowledge, reflecting the inspiration Darwin continues to provide us with. Recall his elegant prose that "There is grandeur in this view of life, . . . from so simple a beginning endless forms most beautiful and most wonderful have been, and are being evolved." The power of his insights continue to motivate research into this "grandeur" view of life as evolutionary biology continues its productive pathway to understanding the biological world. Yet it remains to be fully appreciated what Haeckel highlighted to be the basis of the science of ecology -- " . . . the whole science of the relations of the organism to the environment including, in the broad sense, all the 'conditions of existence.' These are partly organic, partly inorganic …" where he then goes on to note the organic ones include,". . . the entire relations of the organism to all other organisms with which it comes into contact, and of which most contribute either to its advantage or its harm" ( reference; for a modern version see Vandermeer, 1990). As we incorporate more complicated interspecific relations, such as intransitivities and higher order effects into both our theoretical and natural history stories, those "entire relations" are sources of the ecological grandeur exposed in the sublime nature of chaotic complexities (Davis, 2022).

Acknowledgements: NSF grant # DEB-1853261. Insightful comments from Senay Yitbarek aided our discussion of these issues greatly.

**Supplementary Material for: The ghost of ecology in chaos: combining intransitive and higher order effects**

SI: Table S1. Sites and ant species encountered (incompletely reported in Perfecto and Vandermeer, 2020), in three surveys of 25 different coffee farms in Puerto Rico. Note that UTUA 2 is the farm on which the results reported herein were obtained, while UTUA20 is the farm for which the data reported in Vandermeer and Perfecto, 2020, was obtained in an earlier study.

| Site Code | Jan-19 | Jul-19 | Jan-20 | Species Identification |
|---|---|---|---|---|
| UTUA16 | W | W | W | W = *Wasmannia auropunctata* |
| UTUA 2 | W | W | W | S = Solenopsis invicta |
| MARI3 | W | W | W | Tm = Tapinoma melanocephala |
| LASM3 | W | W | W | Mf = Monomorium floricola |
| LASM1 | W | W | W | L = Linepithema iniquum |
| OROC1 | W | W | W | N = Nylanderia pubens |
| UTUA10 | Tm | Tm | Tm | Tb = Tetramorium bicarinatum |
| UTUA20 | S | S | S | C = Cardiocondyla emeryi |
| YAUC3 | S | S | S | Me = Monomorium ebenium |
| UTUA30 | Mf | Mf | Mf | ND = No Dominace |
| ADJU7 | Mf | Mf | Mf | |
| JUAN7 | Mf | Mf | Mf | |
| UTUA18 | L | L | L | |
| MARI2 | W | W | C | |
| JUAN1 | W | Tm | Tm | |
| PONC1 | W | S | S | |
| MARI18 | W | N | W | |
| JAYU3 | Tm | Mf/Tm | Mf | |
| UTUA17 | Tm | Mf | Mf | |
| UTUA5 | S | W/Mf/Tm | W/Mf/Tm | |
| ADJU8 | S | S | Mf | |
| LASM2 | S | Me | W | |
| UTUA13 | S | Mf | S | |
| JAYU2 | ND | Tb | Tb | |

| YAUC4 | ND | ND | ND | |

**SII. The equations for the system** under study, as reported in the text, are:

$$\frac{dS}{dt} = S(1 - S - \alpha_o M - \Psi\gamma_1 P)$$

$$\frac{dP}{dt} = \Psi\gamma_2 PS - mP$$

$$\frac{dM}{dt} = M(1 - M - \alpha_o W)$$

$$\frac{dW}{dt} = W(1 - W - \alpha_{ws}\Theta S)$$

$$\Psi = \frac{1}{1+fS}$$
$$\Theta = \frac{1}{1+\beta P}$$

where S = *Solenopsis*, W = *Monomorium*, W = *Wasmannia*, and P = *Phoridae*. The three competitors (S, M, and W) are in the form of an intransitive loop, symmetrical if $\alpha_0 = \alpha_{ws}$, and, if symmetrical, heteroclinic if $\alpha_i > 2.0$ and dissipative if $\alpha_i < 2.0$. The structure of the basic system is visualizable in figure 1, repeated here.

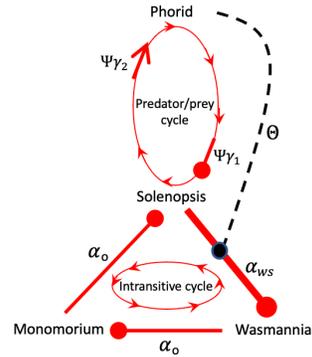

Fixing some of the parameters, we have $\gamma_1 = 1$, $\gamma_2 = 0.6$, m = 0.05. the parameter f =0 for experiments with the classic LV form (without the nonlinear functional response) and f=3 for experiments with the predator/prey system (phorid/Solenopsis) in a limit cycle. The parameters $\alpha_0$ and $\alpha s$ were manipulated to produce either a stable focus ($\alpha_0 < 2$, $\alpha_s$ variable) in the intransitive cycle or a heteroclinic orbit ($\alpha_0 > 2$, $\alpha_s$ variable). The parameter $\beta$ was studied as the basic force for setting the value of $\Theta$.

**SIII: Direct nonlinearity from coupled oscillator** framework: the "symmetrical" case: We begin by setting $\beta = 0$ such that we have the classic case of two coupled oscillators (the intransitive oscillator and the predator prey oscillator). The intransitive loop itself is oscillatory, either dissipating to a single equilibrium point or generating a heteroclinic cycle wherein eventually a single species will predominate, either completely at random or associated with any asymmetry in the system (e.g., if $\alpha_s > \alpha_o$, *Solenopsis* will dominate over the other two species). Here we fix the intransitivity as completely symmetrical ($\alpha_{ws} = \alpha_0$). A summary of all possible, qualitatively distinct, outcomes is presented in table S1.

*Table S1: Summary of qualitatively distinct forms of the model system for the symmetrical case. Each box indicates the starting condition (with predator/prey not connected to the intransitive loop) and the final state of the system (initial "to" final), and, when relevant, the figure associated with the result.*

|  |  | Intransitive loop | | |
|---|---|---|---|---|
|  |  | $\alpha=0.9$ Super-Stable | $\alpha=1.9$ Stable | $\alpha=2.1$ Unstable |
| **Predator prey interaction** | Focal point | Focal point intransitive to 4-spp focal point (Fig S2) | Focal point intransitive to 3-spp focal point, Monomorium extinct (Fig S3) | Heteroclinic intransitive to 3-spp focal point, Monomorium extinct (Fig S4) |
|  | Limit cycle | Focal point to 4-sp Limit cycle (Fig S5) | Focal point to 3-sp limit cycle, Monomorium extinct. | Heteroclinic to 3-sp limit cycle, Monomorium extinct. |

Increasing boundaries of limit cycle

If the uncoupled predatory prey system is a focal point, the final outcome depends on the stability conditions of the intransitive loop, where the equilibrium value of Monomorium declines to a very low level even as the stability of the intransitive loop remains stable. As the intransitive loop enters its independent oscillatory phase, the Monomorium repeatedly descends to a lower level as displayed in figure S2. Further calculations (not shown) make it clear that the oscillations of Monomorium descend to lower and lower levels with the decrease in the "stability" of the intransitive component (i.e. the ai descending), labelled as "increasing boundaries" in Table S1.

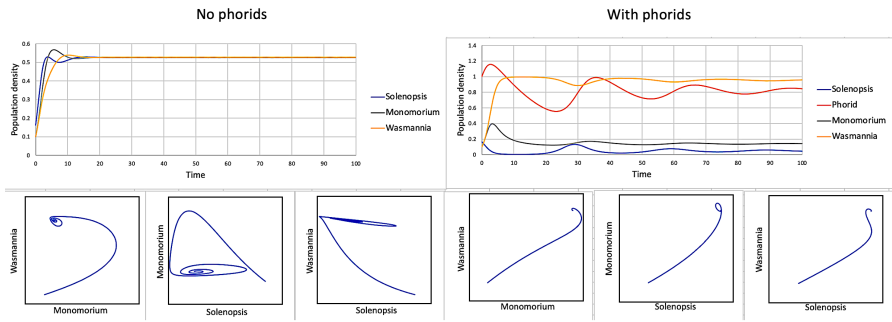

*Figure S2. Change in behavior of the system when initiated at "super-stable" ($a_0 = a_{ws} = 0.9$). a. simple focal point attractor, with no connection to predator. b. resultant stable focus, connection to the predator. Parameters are $\alpha_0 = \alpha_{ws} = 0.9$. $\gamma_1 = \gamma_2 = 1, f = \beta = 0$.*

Even if the intransitive loop is a focal point attractor, if its stability is weak (e.g., a0 = as = 1.9), the addition of the predator effectively drives the equilibrium point of Monomorium toward zero as illustrated in figure S3. Effectively the addition of this higher order effect (the predator/prey system coupled to the intransitive loop) leads to the inevitable extinction of one of the members of the loop, even though there is no direct connection to the predator prey system. Contrary to the general expectations of the literature on predator-mediated coexistence, the action of this predatory system is to drive one of the species to extinction, a consequence of the intransitive/higher-order combination.

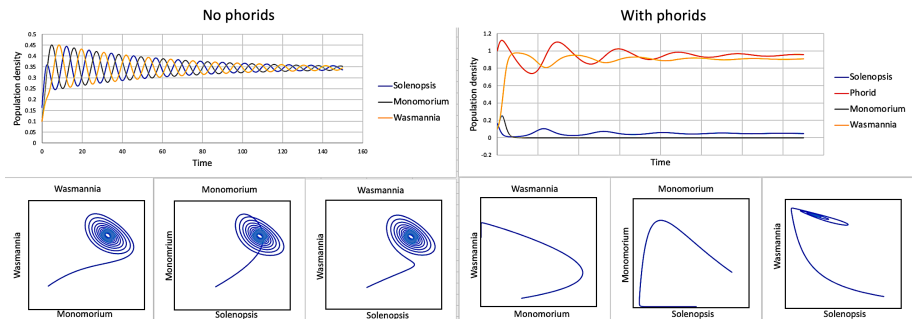

*Figure S3. Change in behavior of the system when initiated at "stable" state (a0 = as = 1.9). a. simple focal point attractor, with no connection to predator. b. resultant stable focus resulting from connection to the predator, leaving a 3-spp focal point attractor with* Monomorium *extinct. Parameters are $\alpha_0 = \alpha_{ws} = 1.9$. $\gamma_1 = \gamma_2 = 1, f = \beta = 0$.*

In the case of a stable focal intransitive system, coupling with a limit cycle predator prey system produces what appears to be a heteroclinic cycle, as in figure S4. Normally a heteroclinic cycle approaches unstable points periodically, approaching those points ever more closely with the passage of time. When the unstable points include zeros, it is the case that which species transgresses a limit at which it will be adjudicated as extinct, is largely a random process. The generalization is that a three species heteroclinic cycle (repeatedly visiting the zero value foreach of the species) will eventually result in the extinction of one species at random. If the cycle is intransitive, that means that the resulting dominant competitors will take over the system. So, in effect, the qualitative outcome of a heteroclinic cycle is a single winner in competition. However, the heteroclinic cycle in this case is driven by the coupling with the predator/prey system and is clearly unbalanced. Thus, which species is most likely to reach the extinction population limit first, is largely decided by the structure of the network, with Monomorium most likely to go extinct first, as illustrated in figure S5. With the extinction of Monomorium, Solenopsis/Phorid system will persist in a limit cycle. For almost all random initiation points this qualitative conclusion holds. However, it is possible to invent initial conditions for which Solenopsis will be driven extinct, in which case the final population in the system will be Wasmannia. For example, the initial conditions S = 0.0000005, P = 1.1738, M = 0.996, W = 0.000185, Solenopsis will be the first to descend below critical limits.

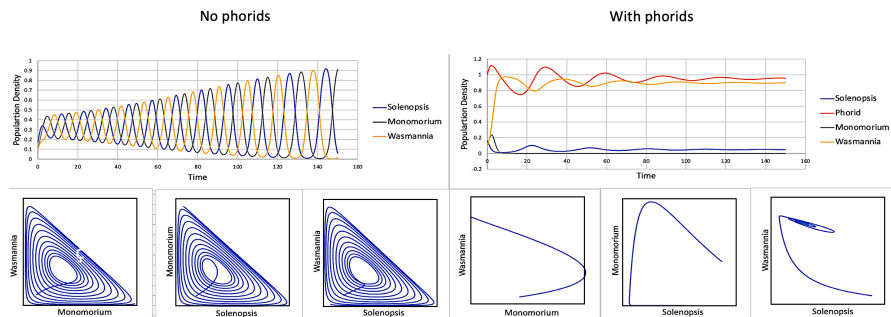

*Figure S4. Change in behavior of the system when initiated at "unstable" state (a0 = as = 2.1). a. heteroclinic cycle, with no connection to predator. b. resultant focal point attractor, with extinction of Monomorium resulting from connection to the limit cycle predator/prey system. Parameters are $\alpha_0 = \alpha_{ws} = 2.1$. $\gamma_1 = \gamma_2 = 1, f = \beta = 0$.*

It is worth noting that in previous work (Vandermeer and Perfecto, 2020), a completely different modeling framework, using a cellular automata approach in a spatially-explicit competition framing, effectively the same result was obtained, if the parameters of the Phorids are properly set. In that case, the natural system indeed did not include *Monomorium* and the reported intransitivity was based on an indirect spatially explicit intransitivity.

Finally, coupling an unstable intransitive loop with the predator/prey in a limit cycle leads to what is effectively a heteroclinic cycle, as shown in figure S5. Yet that cycle has several interesting complications.

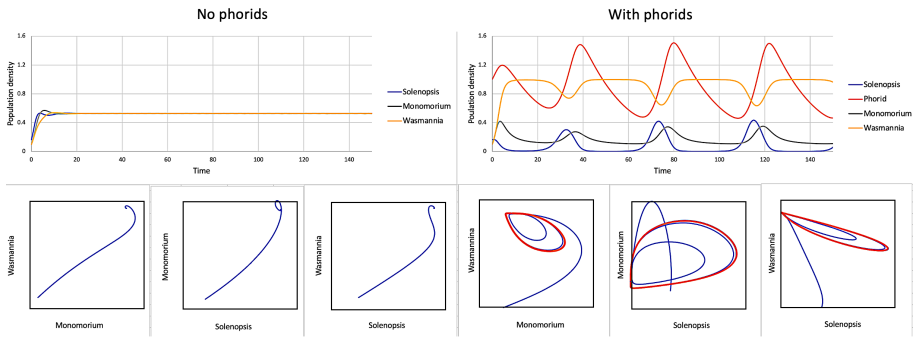

Figure S5. *Change in behavior of the system when initiated at "super stable" state ($\alpha o = \alpha ws = 0.9$). a. simple focal point attractor, with no connection to predator. b. resultant five species limit cycle resulting from connection to the limit cycle predator/prey system.*

All of the above solutions were from the symmetrical intransitive loop ($\alpha_0 = \alpha_{ws}$).

**SIV. Constructing an alternative structure: limits on qualitative interpretation**

A major argument of the text is that both the chaotic attractors and the even more complicated periodic points have a structure that could have been predicted qualitatively from a knowledge of the natural history of the system. That basic structure is that the predator/prey system (the phorid and its prey *Solenopsis*) generates long term oscillations while the connected intransitive oscillations (from *Solenopsis* to *Monomorium* to *Wasmannia*) are more rapid and occur within the time frame of a single oscillation of the predator/prey system. Since the system can be categorized as within the theoretical framing of connected oscillators, why would it not be possible to generate a dynamic in which the intransitive oscillation is long term and the predator/prey oscillation punctuates that long term oscillations with its own rapid oscilations. Searching through parameter space for such a relationship was unsuccessful. The closest we were able to produce is shown in figure S6, where it is evident that the scale of the oscillations is similar, such that a simple chaotic attractor imitating a limit cycle of length 1.0 emerges and all populations have roughly the same time scale of oscillation.

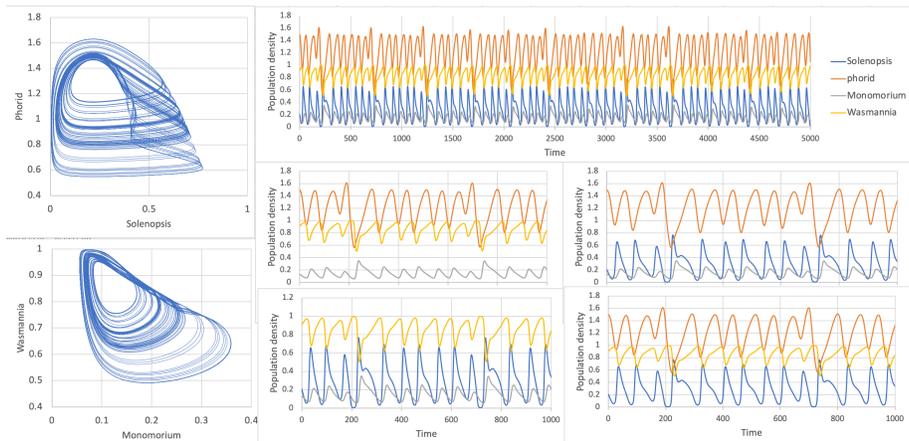

*Figure S6. Chaotic attractor from parameter settings, $\alpha_0 = 1.2$, $\alpha_{ws} = 5$, $\gamma_1 = 1$, $\gamma_2 = 0.6$, $f = 5.45$, $\beta = 2.5$, $m=0.05$.*

29